\documentclass[superscriptaddress,prl,twocolumn,showpacs,preprintnumbers,amsmath,amssymb,bibnotes,secnumarabic,altaffilletter,floatfix]{revtex4}

\usepackage[dvips]{graphicx}
\usepackage{amssymb,amsfonts,amsmath, float}

\begin{document}

\title{Single-Photon Atomic Sorting: Isotope Separation with Maxwell's Demon} 

%\date{\today}
%\author{M. G. Raizen\affil{1}{Center for Nonlinear Dynamics and Department of Physics, The University of Texas at Austin, Austin, TX 78712},
%I. Chavez\affil{1}{},
%\and
%M. Jerkins\affil{1}{}}

%\contributor{Submitted to Proceedings of the National Academy of Sciences
%of the United States of America}

\author{M. Jerkins}
\affiliation{Center for Nonlinear Dynamics and Department of Physics, The University of Texas at Austin, Austin, TX 78712}
\author{I. Chavez}
\affiliation{Center for Nonlinear Dynamics and Department of Physics, The University of Texas at Austin, Austin, TX 78712}
\author{U. Even}
\affiliation{Sackler School of Chemistry, Tel-Aviv University, Tel-Aviv, Israel}
\author{M. G. Raizen}
\affiliation{Center for Nonlinear Dynamics and Department of Physics, The University of Texas at Austin, Austin, TX 78712}

%\maketitle

%\begin{article}
\begin{abstract}
Isotope separation is one of the grand challenges of modern society
and holds great potential for basic science, medicine, energy, and
defense.  We consider here a new and general approach to isotope
separation.  The method is based on an irreversible change of the
mass-to-magnetic moment ratio of a particular isotope in an atomic
beam, followed by a magnetic multipole whose gradients deflect and guide the atoms.  The
underlying mechanism is a reduction of the entropy of the beam by the
information of a single-scattered photon for each atom that is
separated.  We numerically simulate isotope separation for a range of
examples, including lithium, for which we describe the experimental setup we are currently constructing.  Simulations of other examples demonstrate this technique's general applicability to almost the entire periodic table.
We show that the efficiency of the process
is only limited by the available laser power, since one photon on
average enables the separation of one atom.  The practical importance
of the proposed method is that large-scale isotope separation should
be possible, using ordinary inexpensive magnets and the existing technologies of supersonic beams and lasers.
\end{abstract}


\maketitle

%%%%%\section{Introduction \label{sec:intro}}
The efforts to separate isotopes date back to the 1930's and fall into several categories.
Two standard methods of separation are gaseous diffusion and
the ultra-centrifuge~\cite{cent,Diff1,Diff2}.  These methods require many stages of
enrichment and are very inefficient.  Furthermore, these methods are
only suitable for a few elements that can be kept in gas phase, which is a common limitation of isotope separation schemes~\cite{bartlett_patent}.
Isotope separation is also accomplished with mass spectrometry~\cite{MassSep1,MassSep2}.
This method has high isotopic selectivity due to the use of a
quadrupole mass filter, but it is very inefficient due to the low
probability of electron-bombardment ionization and is limited by
space-charge.  In recent years the method of isotope separation by
laser ionization was developed~\cite{LaserSep}.  This approach is highly selective
but requires multiple high-powered lasers for
efficient ionization.  The production rate is also limited by resonant
charge exchange.  

      With this background, it is clear that there is an urgent need for a new and efficient method of isotope separation, the topic of our work.  We first provide an overview of the approach and relate it to the historic problem of Maxwell's Demon.  We then analyze several representative cases using available NIST data~\cite{NIST} and provide the results of numerical simulations.  Finally we discuss the prospects for scalability and experimental realizations.   

%%%%%\section{Single-Photon Atomic Sorting: Method 1}

      We start with a generic prototype for isotope separation: a
      collimated atomic beam of a single element, composed of multiple
      isotopes.  To be more specific, we consider a
      three-level atom with an initial ground state $\mid$i$>$, an
      electronic excited state $\mid$e$>$, and a final state
      $\mid$f$>$.
%\begin{figure}[h]
%\begin{center}
%{\rotatebox{0}{\resizebox{3.4in}{!}{\includegraphics{Figures/schematic.eps}}}}
%\centerline{\includegraphics[width=0.5\textwidth]{Figures/fig1.eps}}
%\caption{A three level atom with ground state $\mid$i$>$. The atom absorbs a photon at wavelength $\lambda$, making a transition to an excited state, $\mid$e$>$, and decays via spontaneous emission to final state $\mid$f$>$. The magnetic moment in the final state is different than the initial state of the atom.\label{fig:levels}}
%\end{center}
%\end{figure}
We further assume that the magnetic moment of state $\mid$i$>$, m$_{\mathrm{i}}$, is
different than the magnetic moment of state $\mid$f$>$, m$_{\mathrm{f}}$.  Now suppose
that an atom crosses a laser beam which induces an irreversible
transition from state $\mid$i$>$ to state $\mid$f$>$ by absorption followed by
spontaneous emission.  The laser
      is tuned to one isotope, changing its magnetic moment, while not
      affecting the others.  The atom then passes through a magnetic
gradient $\nabla$B(x), created by a magnetic multipole. The magnetic multipole acts like a filter, guiding only low-field seeking states.

      We call this process Single-Photon Atomic
      Sorting because each atom is sorted by scattering exactly one
      photon.  It is closely related to a one-way barrier for atoms
      that was used as a general method for cooling the translational
      motion of atoms~\cite{control}.  The goal is to lower
      the entropy of the atomic beam by separating the isotopes.  This process can be viewed as a realization of
      Maxwell's Demon in the sense proposed by Leo Szilard in 1929.
      Here the Demon acts as a sorter, sending each isotope
      in a different direction.  The entropy of the beam cannot be
      lowered with any time-dependent Hamiltonian such as an RF drive~\cite{rf}, and an irreversible step is required.  The atom scatters
      one spontaneous photon from the laser beam, increasing the
      photon's entropy.  This increase compensates for the decrease in the
      entropy of the beam.

\begin{figure}[h]
\begin{center}
%{\rotatebox{0}{\resizebox{3.4in}{!}{\includegraphics{Figures/schematic.eps}}}}
\centerline{\includegraphics[width=0.35\textwidth]{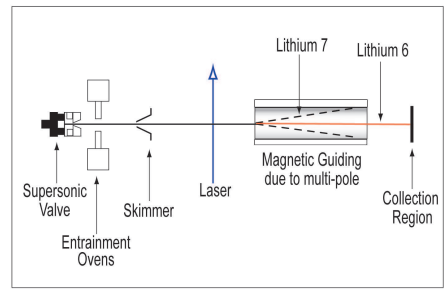}}
\caption{Schematic of the setup for isotope separation of lithium.  Atoms from two ovens are entrained into the flow of a carrier gas from a supersonic nozzle.  A laser is tuned the D$_2$-line of $^7$Li and populates all the atoms into the 2$^2$S$_{1/2}$ F=1 manifold, forcing them to be anti-guided.\label{fig:schematic}}
\end{center}
\end{figure}

      We now discuss a more realistic scenario that we plan to construct in our laboratory.  The starting point
      for this approach must be an atomic beam that has the lowest
      possible entropy of translational motion.  Collimation of an effusive beam is not a viable approach since the resulting flux is too small~\cite{Deflection}.  The best candidate is
      the supersonic beam, which is generated with a high pressure
      carrier gas expanding through a small aperture~\cite{beam}.  Supersonic beams possess remarkable properties, such as an angular
      divergence of only a few degrees and a velocity spread that is
      1$\%$ of the mean velocity~\cite{conden}.  These beams are typically
      pulsed, but for the purpose of isotope separation they should be
      run continuously to maximize throughput.  The desired element
      can be entrained into the flow near the output of the nozzle,
      acquiring the characteristics of the carrier gas.  Efficient
      entrainment can be accomplished using two ovens mounted opposite
      each other and perpendicular to the supersonic flow of atoms,
      as illustrated in Figure~\ref{fig:schematic}. A temperature gradient in each oven can
      produce a collimated effusive beam, which can be enhanced by utilizing a continuous flow reflux design~\cite{refluxOven}.   The ovens are aligned so
      that atoms that are not entrained into the supersonic flow are
      deposited into the opposite oven. This ``atomic ping-pong''
      between ovens greatly reduces the initial amount of material needed
      for separation as well as the background pressure and the need for 
      recycling through vacuum pumps.  Preliminary simulation results show that entrainment efficiencies can be as high as 5-10$\%$ of the initial supersonic beam flux~\cite{Uzi}. Once entrained, the beam
      is collimated with a skimmer and propagates into the laser
      region. The desired isotope will then undergo an irreversible
      change in magnetic moment, differentiating it from the other isotopes
      that are unaffected by the laser beam. After interacting with the laser, atoms proceed to enter a tube surrounded by multipole magnets, which produce a magnetic gradient that guides low-field seeking atoms and anti-guides high-field seeking atoms~\cite{guiding}.

The force due to the
      inhomogeneous magnetic field is
%\begin{equation*}
$F = \mu_{\mathrm{B}}g_{\mathrm{J}}m_{\mathrm{J}}{\nabla}B$,
%\end{equation*}
where $\mu_{\mathrm{B}}$ is the bohr magneton, $g_{\mathrm{J}}$ is the Lande-g factor, $m_{\mathrm{J}}$ is the projection
of the total angular momentum on the quantization axis, and $\nabla$B is the
gradient of the magnetic field. The maximum magnetic field of the
gradient would be strong enough to cause a few elements, such as
lithium, to enter the Paschen-Back regime.  Most of the periodic
table, however, would remain in the weak field limit because of the
strong LS coupling present in heavier atoms. 

\begin{figure}
\begin{center}
%{\rotatebox{0}{\resizebox{3.4in}{!}{\includegraphics{Figures/schematic.eps}}}}
%\centerline{\includegraphics[width=0.4\textwidth]{Figures/Li_rpos_combo.eps}}
\centerline{\includegraphics[width=0.45\textwidth]{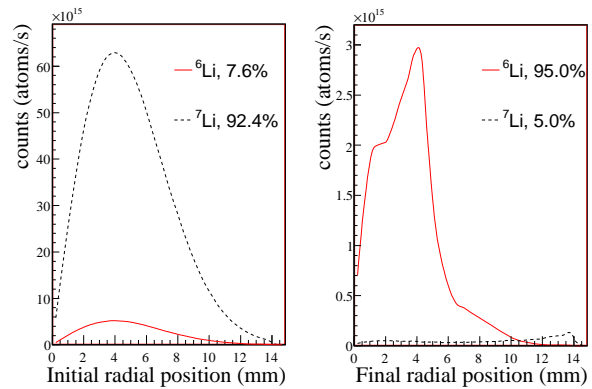}}
\caption{The radial positions of the two lithium isotopes as they enter the magnetic gradient that separates them isotopically, followed by their radial positions upon exiting.\label{fig:Li_rpos}}
\end{center}
\end{figure}

%\section{Lithium}
Lithium serves as a simple example because it has two stable isotopes, $^6$Li and $^7$Li, with natural abundances of 7.6$\%$ and 92.4$\%$ respectively. In our proposed experiment, illustrated in Figure~\ref{fig:schematic}, lithium is entrained into the supersonic beam, and a 670.96~nm laser tuned the $^7$Li D$_2$-line (2$^2$S$_{1/2}$(F=2) $ \rightarrow $ 2$^2$P$_{3/2}$ (F=1 or F=2)) optically pumps the $^7$Li into a high-field seeking state. The laser depletes the $^2$S$_{1/2}$ F=2 manifold  and optically pumps all of the atoms into the $^2$S$_{1/2}$ F=1 manifold. At fields greater than about 50~G, the entire F=1 manifold becomes high-field seeking. This process allows us to efficiently pump all of the unwanted $^7$Li atoms into an anti-guiding mode using a single laser wavelength.  Since we do not excite $^6$Li, we take a statistical loss of one half of the $^6$Li due to the magnetic sub-level projections.  Figure~\ref{fig:Li_rpos} shows the radial distributions of the two lithium isotopes entering the magnetic guiding region, as well as their distributions upon leaving the tube.  The isotope-selective guiding and anti-guiding are clearly evident, and the enrichment can be made arbitrarily high by optimizing the geometry of the tube.  Once an atom collides with the tube walls, the simulation assumes it sticks.  

The results of Figure.~\ref{fig:Li_rpos} were simulated assuming a 0.5~m long region of quadrupole guiding magnets whose specifications are discussed below.  That geometry yields $95\%$ enrichment of $^6$Li, and $36.8\%$ of the $^6$Li that enters the guiding region survives to be collected.  The simulation assumes a beam with a mean velocity of 800~m/s and an initial Gaussian spread of 15~m/s in each component of the beam velocity.  This velocity corresponds to entraining lithium into a beam of helium.  The skimmer shown in Figure~\ref{fig:schematic} is 5~mm in diameter, and approximately $10\%$ of the beam survives it and enters the multipole tube.

%\begin{figure}
%\begin{center}
%{\rotatebox{0}{\resizebox{3.4in}{!}{\includegraphics{Figures/schematic.eps}}}}
%\centerline{\includegraphics[width=0.4\textwidth]{Figures/Li_zpos_hit_wall.eps}}
%\caption{The distance that the unwanted $^7$Li atoms travel down the magnetic quadrupole tube before colliding with the walls, after which the simulation no longer tracks them.\label{fig:Li_zpos}}
%\end{center}
%\end{figure}

\begin{figure}
\begin{center}
%{\rotatebox{0}{\resizebox{3.4in}{!}{\includegraphics{Figures/schematic.eps}}}}
\centerline{\includegraphics[width=0.3\textwidth]{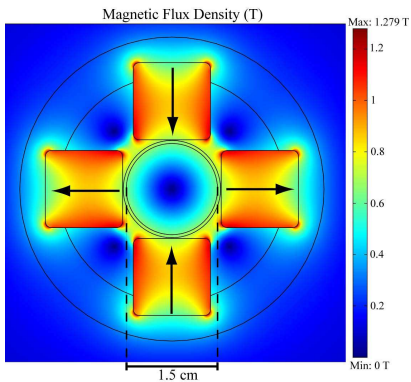}}
\caption{The magnetic flux density of a quadrupole field. The magnets surround a 1.5~cm inner diameter (1.6~cm outer diameter) stainless steel tube and are held in place with an aluminum holder. The magnets are out of vacuum, and the arrows define the direction magnetization.\label{fig:quad}}
\end{center}
\end{figure}

We simulated a quadrupole magnetic field produced by four permanent magnets for the separation of lithium. The magnets are $\frac{1}{2}$'' $\times \frac{1}{2}$'' $\times 1$'' and have a residual flux density of B$_{\rm r}$=1.48~T. The magnets surround a 1.5~cm inner diameter (1.6~cm outer diameter) stainless steel tube.  The resulting magnetic flux density, illustrated in Figure~\ref{fig:quad}, was simulated using finite element analysis. We chose this geometry to avoid putting the magnets in vacuum and to maximize the magnetic field gradients.  While a smaller diameter tube will allow for higher field gradients, it will also reduce the total number of collected atoms of the desired isotope. Although our simulations assume the magnets are held in place with an aluminum holder, slightly higher gradients can be achieved by using a carbon steel holder.

%\section{Metastables}
      One of the distinct advantages of single-photon atomic sorting is that the technique can be applied to almost every atom in the periodic table.  Lithium is particularly easy because it has only two isotopes, and its ground state is $^2$S$_{1/2}$.  The details of the experimental implementation may vary for different elements.  We now discuss two examples that represent qualitatively different categories of elements.  

%\begin{figure}[h]
%\begin{center}
%{\rotatebox{0}{\resizebox{3.4in}{!}{\includegraphics{Figures/schematic.eps}}}}
%\centerline{\includegraphics[width=0.6\textwidth]{Figures/schematic_bend.eps}}
%\caption{Schematic of the set-up for isotope separation for elements with zero magnetic moment in the ground state but a nonzero magnetic moment metastable state that can be excited with a laser. The bend in the magnetic multipole region allows for separation even though the laser only interacts with the desired isotope. \label{fig:schematic_bend}}
%\end{center}
%\end{figure}

\begin{figure}
\begin{center}
%{\rotatebox{0}{\resizebox{3.4in}{!}{\includegraphics{Figures/schematic.eps}}}}
%\centerline{\includegraphics[width=0.4\textwidth]{Figures/Ca_rpos_combo.eps}}
\centerline{\includegraphics[width=0.45\textwidth]{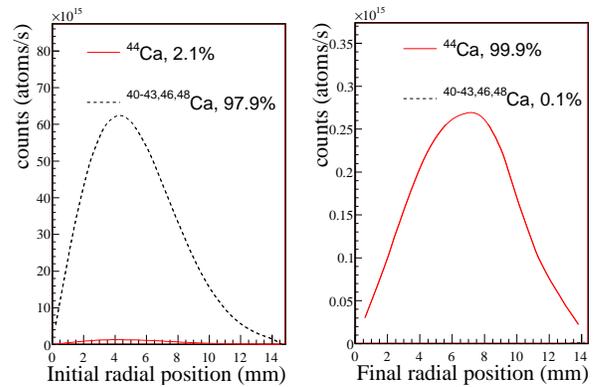}}
\caption{The radial positions of the calcium isotopes as they enter the magnetic gradient that separates them isotopically, followed by their radial positions upon exiting.\label{fig:Ca_rpos}}
\end{center}
\end{figure}

The first example illustrates separation for elements with zero magnetic moment in the ground state and a metastable state that has a nonzero magnetic moment, such as calcium.  By using a 272~nm laser, one can excite $^{44}$Ca to the $^{1}$P$^{\mathrm{o}}_{1}$ state, which quickly decays to the metastable $^{1}$D$_{2}$ state.  Isotopes that are unaffected by the laser will be unaffected by the magnetic gradients since they have zero magnetic moment in the ground state.  The low-field seeking state of the $^{44}$Ca will be guided by the magnetic gradients, meaning that a slight bend in the magnetic guiding tube~\cite{guiding_neutral} will allow the desired isotope to be collected at high enrichment.  Figure~\ref{fig:Ca_rpos} shows simulation results of the enrichment of $^{44}$Ca.  That simulation utilized a 2~m long hexapole magnetic field~\cite{kuiper} created by six of the same magnets described above only arranged around a larger 2.1~cm inner diameter (2.2~cm outer diameter) stainless steel tube.  The tube had a slight bend of 6.0~cm over its 2~m length that served to eliminate the unwanted calcium isotopes.  Since calcium is heavier than lithium, it can be entrained into a heavier metastable gas such as neon, which corresponds to a beam with a mean velocity of 500~m/s and a Gaussian spread of 15~m/s in each component of the supersonic beam velocity.  As shown in Table~\ref{tbl:summary}, our simulated setup collected $9.0\%$ of the $^{44}$Ca that survives the skimmer at $99.9\%$ purity.

\begin{table*}
\centering
%\begin{center}
%\begin{tabular}{|c|c|c|c|c|c|c|c|c|}
\begin{tabular}{|c|c|c|c|c|c|c|}
%%\begin{tabular*}{\hsize}
%%{@{\extracolsep{\fill}}|c|c|c|c|c|c|}

%\begin{tabular}{|>{\centering\arraybackslash}p{0.4in}|>{\centering\arraybackslash}p{0.65in}|>{\centering\arraybackslash}p{0.5in}|>{\centering\arraybackslash}p{0.5in}|>{\centering\arraybackslash}p{0.875in}|>{\centering\arraybackslash}p{0.5in}|>{\centering\arraybackslash}p{0.5in}|>{\centering\arraybackslash}p{0.65in}|>{\centering\arraybackslash}p{0.6in}|}
\hline 
\bf Target& \bf Natural & \bf Laser & \bf Ground & \bf Guiding & \bf Enrich- & \bf Collected\cr
\bf Isotope & \bf Abundance & \bf $\lambda$~(nm) & \bf State & \bf Length (m) & \bf ment &\bf Isotope $\%$\cr
\hline 
$^{6}$Li & 7.6$\%$ & 670.96 & $^2$S$_{1/2}$ & Quad. 0.5 & 95.0$\%$ & 36.8$\%$\cr
\hline 
$^{44}$Ca & 2.1$\%$ & 272.2 & $^{1}$S$_{0}$ & Hex. 2.0 & 99.9$\%$ & 9.0$\%$\cr
\hline
$^{150}$Nd & 5.6$\%$ & 471.9 & $^{5}$I$_{4}$ & Hex. 2.0 & 97.9$\%$ & 23.0$\%$\cr
\hline 
\end{tabular}
\caption{Simulation results of isotope separation from single-photon atomic sorting.\label{tbl:summary}}
\end{table*}

      Many elements, however, do not have a suitable metastable state that
      allows for isotope separation by this method.  We propose an alternate implementation of single photon atomic sorting
      that is much more general.  This method
      will work on any element that has a nonzero magnetic moment in
      the ground state, which includes most of the periodic table.  As the atoms approach the magnetic multipole guiding, a $\sigma^{-}$ polarized laser beam optically pumps our desired isotopes into the stretch low-field seeking state.  Simultaneously a $\sigma^{+}$ polarized laser beam optically pumps the other isotopes into the stretch anti-guided state.  The laser beams can be multi-passed through the supersonic beam until almost all of the atoms have been pumped.  While relying on optical pumping does mean that more than a single photon has to be scattered on average, the isotope separation is accomplished by the scattering of only a small number of photons, which still makes extremely efficient use of the available laser power.  
This method does not rely on a long-lived
	metastable state and is general to all atoms that have a
	ground state magnetic moment, although it does typically require multiple laser wavelengths shifted by a few GHz to optically pump all of the isotopes.  

\begin{figure}
\begin{center}
%{\rotatebox{0}{\resizebox{3.4in}{!}{\includegraphics{Figures/schematic.eps}}}}
%\centerline{\includegraphics[width=0.4\textwidth]{Figures/Nd_rpos_combo.eps}}
\centerline{\includegraphics[width=0.45\textwidth]{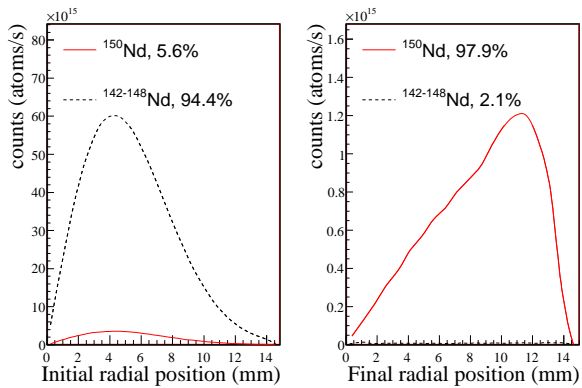}}
\caption{The radial positions of the neodymium isotopes as they enter the magnetic gradient that separates them isotopically, followed by their radial positions upon exiting.\label{fig:Nd_rpos}}
\end{center}
\end{figure}

Figure~\ref{fig:Nd_rpos} shows isotope separation results for a heavier isotope,
$^{150}$Nd, which has a ground state of $^{5}$I$_{4}$.
Using a laser that
promotes the ground state to a J' = 4 excited state, such as a
471.9~nm laser~\cite{Nd}, one could collect 23$\%$ of the $^{150}$Nd that survives the skimmer at 98$\%$ purity.  These simulation results assume a beam with a mean velocity of 500~m/s and a Gaussian spread of 15~m/s in each component of the supersonic beam velocity.  The hexapole magnetic field was a 2~m long tube like the one described for $^{44}$Ca but without the bend.  Similar separation results can be achieved using a 1.8~m long tube with a slight 1~cm bend to aid in eliminating unwanted isotopes.  The precise shape of this bend could be further tuned to achieve the optimal separation geometry.  

The isotope $^{150}$Nd is of
particular interest because it is a double-beta emitter.  Many
experiments are currently investigating neutrinoless double beta decay
in order to determine the neutrino mass and whether neutrinos are
Dirac or Majorana particles~\cite{double}.  SNO+ is one such experiment
currently under development, and it plans to use a large amount of
enriched neodymium to search for neutrinoless double beta decay~\cite{SNO}.
Enriching neodymium is very difficult and can currently only be done
using the atomic vapor laser isotope separation technique~\cite{avlis}.  Hopefully
this simpler approach can aid in the separation of $^{150}$Nd, as well
as other isotopes of interest to physics, medicine, and industry.  

%%%%%\section{Conclusions}
      In conclusion, we have presented single-photon atomic sorting as
      a very general and scalable approach to isotope separation.  The efficiency
      of separation is such that every photon in the laser can provide
      one atom of isotopic interest.  The laser can be recycled in a
      multi-pass configuration until it is depleted.  To put that in
      perspective, a laser with 1 Watt power could separate
      approximately 10$^{19}$ atoms per second, or roughly 500 Moles per
      year.  A supersonic beam can be operated in a continuous mode,
      and the flux is limited only by available vacuum pump speed.
      Diffusion pumps are available with pumping speeds of over
      60,000~L/s, so that large scale separation seems feasible.  The
      next step will be a first experimental demonstration of
      single-photon atomic sorting.  

M.G.R. acknowledges support from the State of
      Texas Advanced Research Program and the Sid W. Richardson
      Foundation.

%\bibliographystyle{apsrev}
%\begin{thebibliography}{999}
%\expandafter\ifx\csname natexlab\endcsname\relax\def\natexlab#1{#1}\fi
%\expandafter\ifx\csname bibnamefont\endcsname\relax
%\def\bibnamefont#1{#1}\fi
%\expandafter\ifx\csname bibfnamefont\endcsname\relax
%\def\bibfnamefont#1{#1}\fi
%\expandafter\ifx\csname citenamefont\endcsname\relax
%\def\citenamefont#1{#1}\fi
%\expandafter\ifx\csname url\endcsname\relax
%\def\url#1{\texttt{#1}}\fi
%\expandafter\ifx\csname urlprefix\endcsname\relax\def\urlprefix{URL
%}\fi
%\providecommand{\bibinfo}[2]{#2}
%\providecommand{\eprint}[2][]{\url{#2}}

%\end{article}
%%%%\bibliography{betadecay_v3}

%%%%\bibliography{betadecay_v3.bbl}

%\end{thebibliography}

\end{document}